\documentclass{article}
\usepackage{spconf}
\usepackage{cite}
\usepackage{amsmath,amssymb,amsfonts}
\usepackage{graphicx}
\usepackage{epstopdf}
\usepackage{subfigure}
\usepackage{stfloats}
\usepackage{textcomp}
\usepackage{xcolor}
\usepackage{flushend}
\usepackage{array}
\usepackage{algorithm}
\usepackage{algpseudocode}
\usepackage{multirow}

\DeclareMathOperator{\sinc}{sinc}

\algnewcommand\algorithmicforeach{\textbf{for each}}
\algdef{S}[FOR]{ForEach}[1]{\algorithmicforeach\ #1\ \algorithmicdo}

\begin{document}

\title{HOW SECURE IS THE TIME-MODULATED ARRAY-ENABLED OFDM DIRECTIONAL MODULATION?}

\name{Zhihao Tao, Zhaoyi Xu, and Athina Petropulu \thanks{This work was supported by ARO grants W911NF2110071, W911NF2320103 and NSF grants ECCS-2033433, ECCS-2320568.}}
\address{Dept. of Electrical and Computer Engineering, Rutgers University, USA}

\setlength{\abovedisplayskip}{3pt}
\setlength{\belowdisplayskip}{3pt}

\maketitle

\begin{abstract}
Time-modulated arrays (TMA) transmitting orthogonal frequency division multiplexing (OFDM) waveforms achieve physical layer security by allowing the signal to reach the legitimate destination undistorted, while making the signal appear scrambled in all other directions. In this paper, we examine how secure the TMA OFDM system is, and show that it is possible for the eavesdropper to defy the scrambling. In particular, we show that, based on the scrambled signal, the eavesdropper can formulate a blind source separation problem and recover data symbols and TMA parameters via independent component analysis (ICA) techniques.  We show how the scaling and permutation ambiguities arising in ICA can be resolved by exploiting the Toeplitz structure of the corresponding mixing matrix, and knowledge of data constellation, OFDM specifics, and the rules for choosing TMA parameters. We also introduce a novel TMA implementation to defend the scrambling against the eavesdropper. 
\end{abstract}


\section{Introduction}
Physical layer security (PLS) \cite{Wyner1975Wire} holds great significance for wireless scenarios, where traditional cryptographic methods may fail to provide low latency and scalability
\cite{poor2017wireless}. 
%
%
A recently introduced way to achieve PLS is directional modulation (DM), which preserves the information signals only along the pre-selected legitimate directions, while distorting them along all other spatial directions \cite{Daly2009DM, su2021secure,nooraiepour2022space}.
Different from  PLS approaches such as cooperative relaying strategies~\cite{dong2009improving,Li2020relay,Li2011cooperative} and artificial noise~\cite{zhang2019AN,wang2017AN}, DM operates without the need for channel state information, and without generating interference to the legitimate receiver.
%
%
One way to implement DM is by operating on the signal to be transmitted \cite{Daly2009DM, Kalantari2016DM,Alodeh2016DM}, or  
%
 by manipulating the transmitter hardware \cite{ding2017free,ding2017Circular,Alotaibi2016subset,Ding2019TMA_OFDM}. In \cite{Ding2019TMA_OFDM}, 
 Time-modulated arrays (TMA) using orthogonal frequency division multiplexing (OFDM) waveforms are proposed, wherein each transmit antenna operates in a periodic ON-OFF pattern. This  type of operation generates harmonic signals at the OFDM subcarrier frequencies. Through careful design of the ON-OFF pattern, it is possible to receive the information signals undisturbed in specific directions, while they appear scrambled in all other directions. The scrambling occurs because the symbols of each subcarrier are mixed with the harmonic signals from all other subcarriers. 
As compared to other DM arrays, the TMA  transmitter of \cite{Ding2019TMA_OFDM} has the capability to alter both the magnitude and phase of the transmitted signal while utilizing only one radio-frequency (RF) chain, resulting in reduced hardware expenses.  


Previous studies on the TMA DM technique have focused on hardware implementation, energy efficiency improvement, ON-OFF pattern designs and its applications \cite{Ding2019TMA_OFDM, nooraiepour2022time, Purushothama2023Synthesis, Guo2018time, li2022chaotic, Huang2021multi, Huang2022target,xu2023TMA}, and have not studied how secure the TMA DM system is. An exception is the work in \cite{nooraiepour2022time}, which examines whether the eavesdropper can use machine learning techniques to estimate the parameters of the OFDM waveform and then spoof the receiver using a similar waveform. The conclusion of \cite{nooraiepour2022time} was that DM can prevent such spoofing. 

In this paper, we investigate the level of security provided by the TMA achieved scrambling,  and show that, unless certain action is taken, the TMA OFDM system is actually not secure enough. In particular, we show that for a typical TMA OFDM transmitter \cite{Ding2019TMA_OFDM}, the signals received by the eavesdropper on all subcarriers can be used to formulate a linear system of the form $\boldsymbol{y}=\boldsymbol{V} \boldsymbol{s}$, where ${\bf \boldsymbol{V}}$  is the mixing matrix that depends on the TMA parameters and the legitimate user direction, and $\boldsymbol{s}$ contains the information symbols. ${\bf \boldsymbol{V}}$ is of course unknown to the eavesdropper, however, under certain conditions, classical blind source separation methods, such as independent component analysis (ICA), can aid the eavesdropper in estimating it within some ambiguities. The conditions include statistical independence and non-Gaussianity of the source signals (or at most one can be Gaussian) \cite{Aapo2000ica}. The ambiguities include scaling of the columns of ${\bf \boldsymbol{V}}$, and column order ambiguity. We show that these ambiguities can be resolved by exploiting the  Toeplitz structure of the corresponding mixing matrix, and knowledge of the transmission constellation, the OFDM parameters and the rules for choosing the TMA parameters. We should note that knowing the TMA parameter assignment conditions does not mean that the specific TMA parameters are known. Upon resolving the ambiguities the data symbols and the TMA parameters can be recovered. 
We also propose a  TMA implementation mechanism to defend the scrambling against the eavesdropper by varying the TMA ON-OFF pattern randomly over time, thus degrading the ICA applicability. Complexity analyses and numerical experiments demonstrate that our proposed defying and defending approaches are effective and efficient.

\vspace*{-3mm}
\section{System model}
\vspace*{-2mm}
We consider a single RF-chain  TMA-enabled OFDM DM transmitter as proposed in \cite{Ding2019TMA_OFDM}, comprising a uniform linear array with $N$ elements spaced by half wavelength, and OFDM waveforms with $K$ subcarriers spaced by $f_s$. Here we do not consider power allocation at the transmitter end or noise at the receiver end; the power of each antenna in each subcarrier is set to be identical. Let $s_k$ be the digitally modulated data symbol assigned to the $k$-th subcarrier. The  OFDM symbol equals
${x}(t) = {1}/{\sqrt{K}} \sum \limits_{k=1}^K s_k \exp \{{j2 \pi [f_0+(k-1)f_s]t\}},$
where $f_0$ denotes the frequency of the first subcarrier and $1/\sqrt{K}$ is the power normalization coefficient that normalizes $s_k$ to be unit power. Note that we eliminate the index of the transmitted OFDM symbol here as it is appropriate to consider only one OFDM symbol in the following analyses.

The OFDM symbol radiated 
towards direction $\theta \in [0, \pi]$ can be expressed as 
\begin{equation}
{y}(t, \theta) = \frac{1}{\sqrt{N}} \sum \limits_{n=1}^N {x}(t)  w_n  U_n(t)  e^{j(n-1) \pi \cos \theta},
\end{equation}
where 
$w_n$ is the $n$-th antenna weight, and $U_n(t)$ represents the ON-OFF  switching function of the $n$-th antenna. 
In order to focus the beam towards the direction of the legitimate user,  $\theta_0$, we set $w_n = e^{-j(n-1) \pi \cos \theta _{\rm 0}}$. The ON-OFF switching function $U_n(t)$ is usually designed as a square waveform as in \cite{Ding2019TMA_OFDM}. Let the normalized switch ON 
time instant and the normalized ON time duration be respectively denoted by $\tau_n^o$ and $\Delta \tau_n$. $U_n(t)$ can be expressed in Fourier series as the sum of complex sinusoids with  frequencies  $mf_s$ and corresponding coefficients
$a_{mn}\mkern-4mu =\mkern-4mu \Delta \tau _n \sinc (m \pi \Delta \tau _n)  e^{-jm \pi (2 \tau _n^o + \Delta \tau _n)}$, where $m$ is an integer, and $\sinc (\cdot)$ is an unnormalized sinc function. By combining the above equations, we  write the transmitted symbol as
\begin{equation*}\label{eq5}
    {y}(t, \theta) = \frac{1}{\sqrt{NK}}  \sum_{k=1}^K s_k  e^{j2 \pi [f_0+(k-1)f_s]t}  \mkern-12mu\sum_{m=-\infty}^{\infty} \mkern-12mu e^{j2m \pi f_s t}V_m,
\end{equation*}
where
\begin{equation}\label{eq7}
    V_m =  \sum \limits_{n=1}^N a_{mn} e^{j(n-1) \pi (\cos \theta - \cos \theta _{\rm 0})}.
\end{equation}

The signal seen in direction $\theta$ on the  $i$-th subcarrier equals
\begin{equation}\label{eq8}
    {y}_i (t, \theta) = \frac{1}{\sqrt{NK}} \sum \limits_{k=1}^K s_k e^{j2 \pi [f_0+(k-1)f_s]t} V_{i-k}.
\end{equation}
After OFDM demodulation, the scrambled data symbol in \eqref{eq8} can be expressed as $y_i (\theta) = 1/\sqrt{NK} \sum_{k=1}^K s_k V_{i-k}$.
In order to achieve  DM functionality, treating only $\theta_0$ as the legitimate use direction,   $\tau _n^o$ and $\Delta \tau _n$ are chosen  to satisfy $V_{m \ne 0}(\tau _n^o,\Delta \tau _n,\theta = \theta _0) = 0$ and $V_{m = 0}(\tau _n^o,\Delta \tau _n, \theta = \theta _0) \ne 0$, which can be achieved by  the following three conditions \cite{Ding2019TMA_OFDM}: (C1) $\Delta \tau _n \in [0, 1], \tau _n^o \in \{\frac{h-1}{N}\}_{h=1,2,...,N}$ (note that the subscript $n$ is not necessarily equal to $h$); (C2) $\tau _p^o \ne \tau _q^o, \Delta \tau _p = \Delta \tau _q$ for $p \ne q$; and (C3) $\sum_{n=1}^N \Delta \tau _n \ne 0$. 
At the receiving end,  the received OFDM signal along the legitimate use $\theta _0$ equals ${y}(t, \theta _0) = \Delta \tau _n \sqrt{N/K} {s}(t)$, while in all other directions, the signal is scrambled.

\vspace{-3mm}
\section{Defy and defend the scrambling}
\vspace{-1mm}

Let us assume an eavesdropper in direction $\theta$. The demodulated signals on all 
subcarriers at the eavesdropper,  put in vector $\boldsymbol{y}$
can be expressed as  
\begin{equation}
    \boldsymbol{y} = \boldsymbol{V} \boldsymbol{s}, \label{BSS}
\end{equation} where  $\boldsymbol{V} \in \mathbb{C}^{K\times K}$ is a Toeplitz, whose $(i,j)$ element equals
\begin{equation}\label{eq9}
\boldsymbol{V}(i,j) = \frac{1}{\sqrt{NK}} V_{i-j},\quad i,j = 1,2,...K,
\end{equation}
and $\boldsymbol{s} = [s_1, s_2, \cdots, s_K]^T$. 
One can see that, due to (C1)-(C3),  along direction  $\theta \ne \theta _0$, the received signal during one OFDM symbol is scrambled by the data symbols modulated onto all other subcarriers since the mixing matrix $\boldsymbol{V}$ is no longer diagonal. 

In (\ref{BSS}), the elements of $\boldsymbol{s}$ are 
symbols transmitted on different subcarriers,  which are statistically independent and non-Gaussian (they are usually uniformly distributed).
 Therefore, the  eavesdropper can leverage  $H$ received OFDM symbols
 to estimate $\boldsymbol{V}$  and recover the  transmitted data symbols via an 
 ICA method. 

\smallskip
\noindent\textit{Assumptions - } The assumptions made here are  as follows. We assume that the eavesdropper knows the OFDM specifics of the transmitted signals, like the number of subcarriers, $K$, the data modulation scheme, and the above-stated rules  (C1)-(C3) for implementing TMA. Note that (C1)-(C3) define a set of values for the TMA parameters so knowing the rules does not mean that the eavesdropper knows the specific TMA parameters used at the transmitter. 

ICA attempts to decompose the linearly mixing data based on the {independence} assumption on source signals and the important fact that the sum of two or more independent random variables is more Gaussian than the original variables. Mathematically, ICA tries to find an unmixing matrix $\boldsymbol{W}$, so that
the elements of $\boldsymbol{W} \boldsymbol{y}$ is as non-Gaussian as possible.
%
In this work, we adopt a quantitative index, i.e., negentropy, to measure non-Gaussianity and implement
FastICA \cite{Aapo1999fast} to find $\boldsymbol{W}$ in a very fast and reliable fashion.

\begin{algorithm}[t]
    \caption{Reordering Algorithm}\label{reorder}
    \begin{algorithmic}[1]
        \State Calculate the amplitude of each elements in $\boldsymbol{F}$ and get a new matrix $\boldsymbol{Q}$, the $i$th column of which is denoted by $\boldsymbol{q}_i$;
        \ForEach {$i = 1,2,...,K$}
            \State Take $\boldsymbol{q}_i(1)$ as the first diagonal element in the first row of $\boldsymbol{Q}$;
            \State Find the closest elements to $\boldsymbol{q}_i(1)$ in the remaining rows of $\boldsymbol{Q}$ and put them in the corresponding diagonal placements;
            \State Obtain a diagonal vector $\boldsymbol{d}$ after step 4 and normalize it by $\boldsymbol{d} / \|\boldsymbol{d}\|$;
            \State Compute the standard deviation $\sigma _i$ of normalized $\boldsymbol{d}$;
        \EndFor
         
        \State Let $\boldsymbol{\sigma} = [\sigma _1, ..., \sigma _K]$ and find the index of the minimum element in $\boldsymbol{\sigma}$ as $I$;
        \State Let $i=I$ and execute steps 3 and 4, we can obtain a reordered $\boldsymbol{Q}$ and accordingly reordered $\boldsymbol{F}$.
    \end{algorithmic}
    \vspace*{-0.5mm}
\end{algorithm}

\begin{algorithm}
    \caption{Phase Ambiguity Resolving Algorithm}\label{phase}
    \begin{algorithmic}[1]
        \State Obtain $\{\boldsymbol{F}_u \}_{u=1,2,...,M}$ according to the transmission constellation and the Toeplitz structure;
        \State Calculate the ratio of the real part and the imaginary part of each $\boldsymbol{F}_u$, denoted as $\{\lambda _u \}_{u=1,2,...,M}$, respectively;
        \ForEach {$\lambda _u$}
            \State Compute $N_u$ and $\Delta \tau _u$ according to (\ref{Eq1}) and $ \lambda = 1/{\tan({\frac{N-1}{2} \pi \varphi})}$;
            \State Check if $N_u \in \mathcal{G}_{N}$ and if $\Delta \tau _u \in [0,1]$: if both are yes, keep this group of solutions; otherwise, discard them;
        \EndFor
        \If {Only one group of $N_u$ and $\Delta \tau _u$ found}
            \State \textbf{Return} $\boldsymbol{F}_u$ corresponding to this group of solutions;
        \Else
            \ForEach {group of $N_u$ and $\Delta \tau _u$}
                \State Check if $\{\tau _n^o\}_{n=1,2,...,N}$ can be found by (C1)-(C3) and (\ref{eq9}): if yes, keep this group of solutions and return the corresponding $\boldsymbol{F}_u$; otherwise, discard them.
            \EndFor
        \EndIf
    \end{algorithmic}
\end{algorithm}

\subsection{On resolving ambiguities} \label{3.1}
The $\boldsymbol{W}$ obtained by ICA, contains scaling and permutation ambiguities, which would prevent recovery of the data symbols.
To resolve those ambiguities We will resort to prior knowledge, as reflected in our assumptions.


%
The scaling ambiguity can be divided into amplitude and phase ambiguity.  The amplitude ambiguity arises because a data symbol can be scaled by a real-valued constant and the corresponding column of the mixing matrix by the inverse constant, without affecting the observation vector.  ICA  cannot distinguish between the data symbols and the scaled ones as they both have the same level of non-Gaussianity. However, by knowledge of the transmit constellation, the amplitudes of source signals are known and hence one can use this information to determine the amplitude scaling.
%


The permutation ambiguity derives from the fact that the order of elements in $\boldsymbol{s}$ and the order of each column in $\boldsymbol{V}$ can change correspondingly and this will not change  $\boldsymbol{y}$. Therefore, ICA cannot identify the data symbols in the recovered vector $\boldsymbol{W}\boldsymbol{y}$ in the right order, in other words, it cannot tell which data symbol is assigned to which subcarrier. To solve this issue, we proceed as follows. We define $\boldsymbol{F}\buildrel \triangle \over = \boldsymbol{W} ^{-1}$. In the absence of ambiguities, $\boldsymbol{F}$ would have been equal to $\boldsymbol{V}$ and thus a Toeplitz matrix.
We propose to reorder $\boldsymbol{F}$, trying to see whether the reordering creates a Toeplitz matrix. Considering that there are $K!$  possible orderings, and that the main diagonal elements can determine the Toeplitz structure of $\boldsymbol{F}$, we focus on the main diagonal elements to reduce the computation time. We  use standard deviation, $\sigma$, to measure the similarity of the main diagonal elements (note that there are small computational errors within ICA and hence the estimated diagonal elements are usually not the same). We summarize the proposed reordering procedures in Algorithm \ref{reorder}, the complexity of which is $O(K^3)$.
%

When the scalar is complex,   the scaling ambiguity will also introduce a phase ambiguity.
We can use the Toeplitz structure to reduce the phase uncertainty first. Consider $M$-PSK modulation for example (the extensions to QAM modulation are straightforward as QAM can be viewed as the combination of several kinds of PSK modulation). There will be $M^K$ phase possibilities for $\boldsymbol{F}$ and the Toeplitz constraint can reduce it to $M$. This is because the phases of diagonal elements of $\boldsymbol{F}$ must be the same and each source signal can have up to $M$ phase transformations. We define these $M$ possibilities for $\boldsymbol{F}$ as $\boldsymbol{F}_1, \boldsymbol{F}_2, ..., \boldsymbol{F}_M$. To determine which one is the actual mixing matrix we proceed as follows.

Let $\Delta \tau _n = \Delta \tau $, $\varphi = \cos \theta - \cos \theta _{\rm 0}$. From (\ref{eq7}) we get that
\begin{equation}\label{Eq1}
\begin{split}
    V_0 & = \Delta \tau \sum \limits_{n=1}^N e^{j(n-1) \pi \varphi} 
     = 
    \Delta \tau \frac{\sin(\frac{N}{2} \pi \varphi)}{\sin(\frac{1}{2} \pi \varphi)} e^{j\frac{(N-1)}{2} \pi \varphi}.
\end{split}
\end{equation}
The ratio of the real part and imaginary part of $V_0$ is $ \lambda = 1/{\tan({\frac{N-1}{2} \pi \varphi})}$
which is also the ratio of the real part and imaginary part of $\boldsymbol{V}(1,1)$. To resolve the remaining phase uncertainty, 
%
%
We check whether  there exist solutions of $N$, $\Delta \tau$, $\{\tau _n^o\}_{n=1,2,...,N}$ according with (C1)-(C3) and $\varphi$ that correspond to exactly one of the elements of $\{\boldsymbol{F}_u \}_{u=1,2,...,M}$. Let us denote the range of possible values for  $N$ as $\mathcal{G}_{N}$, and set the maximum value in $\mathcal{G}_{N}$ larger than $N$.
%

On assuming that $\varphi$ is known, which can be obtained by direction of arrival algorithms, the steps of resolving the phase ambiguity are exhibited in Algorithm \ref{phase}. This proposed algorithm can always find only one group 
of $N$, $\Delta \tau$ and $\{\tau _n^o\}_{n=1,2,...,N}$ that corresponds to one of $\{\boldsymbol{F}_u \}_{u=1,2,...,M}$ and this one is what we are looking for. The maximum complexity of Algorithm \ref{phase} is $N!$ due to $\tau _n^o \in \{\frac{h-1}{N}\}_{h=1,2,...,N}$. The principles behind Algorithm \ref{phase} are the following.  Except in some specific cases where $\varphi = 1/J$ ($J=\pm 1, \pm 2, \pm 3, ...$), there is only one 
$N$ that can be found for each $\lambda _u$.
Assuming that two $N$ are found corresponding to the same $\lambda_u$,
we will get that $\frac{N_1}{2} \pi \varphi + L \pi = \frac{N_2}{2} \pi \varphi$ ($N_1 \neq N_2, L=\pm 1, \pm 2, \pm 3, ...$) and hence $N_2 - N_1 = 2L / \varphi$. This condition is rare to be satisfied since the value of $N$ and the precision of $\theta$ are usually limited in practice. 
Then, by further exploiting the rules (C1)-(C3) for $\Delta \tau$ and $\tau _n^o$, this algorithm will exclude all incorrect solutions and the remaining ones will be the actual parameters of the mixing matrix. In that way,  we eliminate the phase ambiguity. Even when $\varphi$ is not known, Algorithm \ref{phase} can work well; the only change is that we need to search for $\varphi_u$ and $N_u$, $\Delta \tau _u$ simultaneously. Overall, the maximum complexity of our proposed defying scheme is $O(ICA)+O(K^3+N!)$, where $O(ICA)$ can be neglected as FastICA usually runs very fast.
\vspace*{-3mm}

\linespread{1.16}

{\small 
\begin{table}[t]
    \begin{center}
    \caption{Average BER of the TMA OFDM DM system}\label{T1}
    \renewcommand{\arraystretch}{0.85}
    \begin{tabular}{p{0.40cm}<{\centering}|p{0.7cm}<{\centering}|p{0.7cm}<{\centering}|p{1.1cm}<{\centering}|p{0.85cm}<{\centering}|p{0.85cm}<{\centering}|p{0.85cm}<{\centering}}
    \hline
    \hline
    No. & $\theta _0$($^\circ$) & $\theta _e$($^\circ$) & $\varphi$ & BER1 & BER2 & BER3   \\ \hline
    1   & 50    & 90         & -0.6428   & 0.3080      & 0      & 0.4504    \\ 
    2   & 60    & 30         & 0.3660    & 0.2640      & 0      & 0.5218    \\ 
    3   & 80    & 40         & 0.5924    & 0.4474      & 0      & 0.5004    \\ \hline
    4   & 30    & 70         & /         & 0.5487      & 0      & 0.4168    \\ 
    5   & 40    & 90         & /         & 0.3754      & 0      & 0.4824    \\ 
    6   & 50    & 130        & /         & 0.2744      & 0      & 0.4789    \\ \hline
    \end{tabular}
    \label{TABLE1}
    \end{center}
    \vspace*{-7mm}
\end{table}
\linespread{1.16}
}

\vspace*{-2mm}
\subsection{Proposed Defending Mechanism} \label{3.2}
\vspace*{-2mm}
Since the above ICA can work only in stationary environments and necessitates long data for estimating the required higher-order statistics,   we can disturb the applicability of ICA by changing the mixing matrix of TMA over time.  This can be done by selecting randomly $\{\tau _n^o\}_{n=1,2,...,N}$ in each OFDM symbol period according to $\tau _n^o \in \{\frac{h-1}{N}\}_{h=1,2,...,N}$ and $\tau _p^o \ne \tau _q^o$. Also, this mechanism is able to maintain the DM functionality as it still satisfies the scrambling scheme described in Section II.

\vspace*{-2.1mm}
\section{NUMERICAL RESULTS}
\vspace*{-2mm}
We present numerical results to evaluate our proposed scrambling defying approach and defending mechanism.

We simulated a TMA OFDM scenario with parameters  $N = 7$, $K = 16$, ${H} = 1$e5 and for the data  we used BPSK modulation. We conducted $6$ experiments and for each experiment the direction of the legitimate user, $\theta _0$,  and the eavesdropper, $\theta _e$, were chosen differently, as shown in Table \ref{T1}. In the experiments No. $1-3$, $\varphi$ is set as known, while in experiment No. $4-6$ it is unknown. Denote the bit error rate (BER) at $\theta _e$ based on the raw signal received by the eavesdropper as BER1, based on the signal recovered via ICA and the proposed ambiguity resolving algorithms in Section \ref{3.1} as BER2, and based on the signal recovered via ICA and ambiguity resolving after the proposed defending mechanism in Section \ref{3.2} was applied at the transmitter as BER3. For each experiment, we generated randomly 30 different groups of $\Delta \tau = 1/N$ and $\{\tau _n^o\}_{n=1,2,...,N} = (n-1)/N$ according to the rules (C1)-(C3) to compute the average BER1, BER2 and BER3. The results are shown in Table \ref{T1},
from which we can find that, in all cases, the BER1 is not $0$, representing the effect of scrambling at the eavesdropper. In all cases, the  BER2 is $0$, indicating that the eavesdropper is able to defy scrambling and recover the source symbols. Meanwhile, in all cases, the BER3 is not $0$, despite the eavesdropper trying to defy scrambling, which demonstrates that the proposed mechanism is effective in defending the scrambling.

\begin{figure}[t]
\centerline{\includegraphics[width=2.5in]{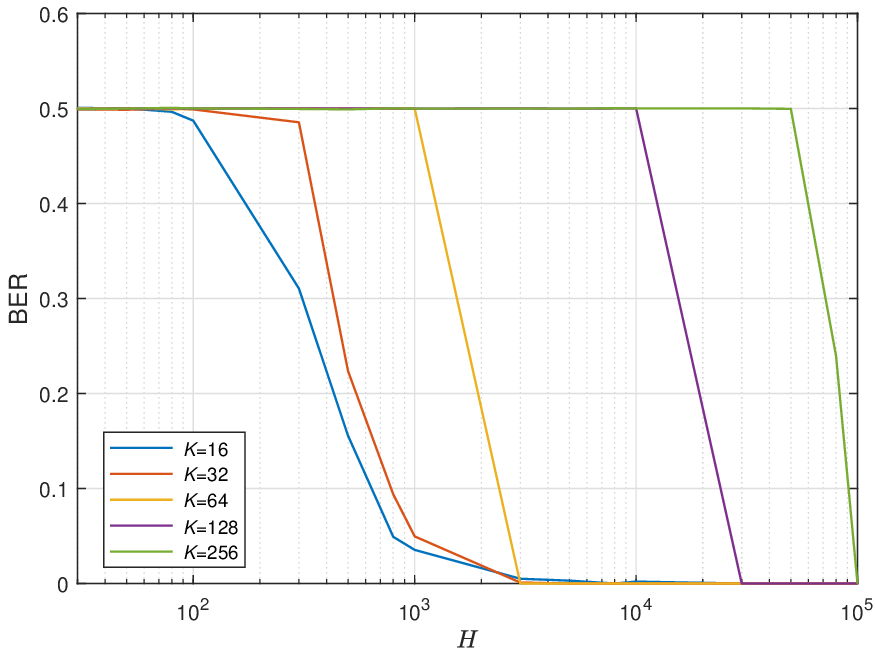}}
\vspace*{-2mm}
\caption{\small Defying performance of the proposed scheme with respect to different $K$ and ${H}$.
}\label{fig1}
\vspace*{-5mm}
\end{figure}

Fig. \ref{fig1} shows the BER  of the proposed defying scheme with respect to different numbers of OFDM subcarriers, $K$ and OFDM symbols, $H$. In this figure we set $\theta _0=60^{\circ}$ and $\theta _e=30^{\circ}$, $N = 7$, $\Delta \tau = 1/N$, $\{\tau _n^o\}_{n=1,2,...,N} = (n-1)/N$, $\varphi$ is taken  known and adopt BPSK modulation. From Fig. \ref{fig1}, we observe that the BER can be reduced to 0 even when $K=256$, showcasing the great potency of our proposed defying scheme. 
As expected, the defying performance improves with ${H}$ as the ICA performance improves. Moreover, it can be seen that ICA will require many more samples to make a complete defying when $K$ is large. This is because a larger $K$ means a larger number of source signals, and thus   ICA needs more samples to work well. Additionally, when  $K$ is large, $V_m$, for $|m|$  close to $K$, in $\boldsymbol{V}$ will be very small due to the term $\sinc (m \pi \Delta \tau _n)$ in (\ref{eq7}). Considering there are also some small estimation errors in ICA, $|V_m|$, for $K$ large and $|m|$ close to $K$,  will be even smaller than the estimation errors of ICA, which will eventually lead to failure of Algorithm \ref{reorder} to reorder accurately. Therefore, a large number of samples are needed to improve the accuracy of ICA estimates and accordingly the performance of the reordering algorithm.

\vspace*{-1.4mm}
\section{CONCLUSION}

In this paper, we have validated that the data scrambling of TMA OFDM DM systems can be defied by an eavesdropper; this has never been studied in the existing literature. We have introduced a novel ICA-based defying scheme that can infer the mixing matrix with no ambiguities. Also, we have proposed a simple TMA implementation mechanism to defend the defying of ICA. Numerical results have demonstrated the effectiveness and efficiency of proposed defying and defending approaches. 

\vfill\pagebreak

\bibliography{ConfPaper_2023}

\begin{thebibliography}{10}

\bibitem{Wyner1975Wire}
A.~D. Wyner,
\newblock ``The wire-tap channel,''
\newblock {\em The Bell System Technical Journal}, vol. 54, no. 8, pp. 1355--1387, 1975.

\bibitem{poor2017wireless}
H.~V. Poor and R.~F. Schaefer,
\newblock ``Wireless physical layer security,''
\newblock {\em Proceedings of the National Academy of Sciences}, vol. 114, no. 1, pp. 19--26, 2017.

\bibitem{Daly2009DM}
M.~P. Daly and J.~T. Bernhard,
\newblock ``Directional modulation technique for phased arrays,''
\newblock {\em IEEE Transactions on Antennas and Propagation}, 2009.

\bibitem{su2021secure}
N. Su, F. Liu, and C. Masouros,
\newblock ``Secure radar-communication systems with malicious targets: Integrating radar, communications and jamming functionalities,''
\newblock {\em IEEE Transactions on Wireless Communications}, vol. 20, no. 1, pp. 83--95, 2021.

\bibitem{nooraiepour2022space}
A. Nooraiepour, S. Vosoughitabar, C.-T.~M. Wu, W.~U. Bajwa, and N.~B. Mandayam,
\newblock ``Programming wireless security through learning-aided spatiotemporal digital coding metamaterial antenna,''
\newblock {\em Advanced Intelligent Systems}, 2023.

\bibitem{dong2009improving}
L. Dong, Z. Han, A.~P. Petropulu, and H.~V. Poor,
\newblock ``Improving wireless physical layer security via cooperating relays,''
\newblock {\em IEEE Transactions on Signal Processing}, 2010.

\bibitem{Li2020relay}
Q. Li and L. Yang,
\newblock ``Beamforming for cooperative secure transmission in cognitive two-way relay networks,''
\newblock {\em IEEE Transactions on Information Forensics and Security}, vol. 15, pp. 130--143, 2020.

\bibitem{Li2011cooperative}
J. Li, A.~P. Petropulu, and S. Weber,
\newblock ``On cooperative relaying schemes for wireless physical layer security,''
\newblock {\em IEEE Transactions on Signal Processing}, vol. 59, no. 10, pp. 4985--4997, 2011.

\bibitem{zhang2019AN}
W. Zhang, J. Chen, Y. Kuo, and Y. Zhou,
\newblock ``Artificial-noise-aided optimal beamforming in layered physical layer security,''
\newblock {\em IEEE Communications Letters}, 2019.

\bibitem{wang2017AN}
W. Wang, K.~C. Teh, and K.~H. Li,
\newblock ``Artificial noise aided physical layer security in multi-antenna small-cell networks,''
\newblock {\em IEEE Transactions on Information Forensics and Security}, vol. 12, no. 6, pp. 1470--1482, 2017.

\bibitem{Kalantari2016DM}
A. Kalantari, M. Soltanalian, S. Maleki, S. Chatzinotas, and B. Ottersten,
\newblock ``Directional modulation via symbol-level precoding: {A} way to enhance security,''
\newblock {\em IEEE Journal of Selected Topics in Signal Processing}, 2016.

\bibitem{Alodeh2016DM}
M. Alodeh, S. Chatzinotas, and B. Ottersten,
\newblock ``Energy-efficient symbol-level precoding in multiuser {MISO} based on relaxed detection region,''
\newblock {\em IEEE Transactions on Wireless Communications}, vol. 15, no. 5, 2016.

\bibitem{ding2017free}
Y. Ding and V. Fusco,
\newblock ``A synthesis-free directional modulation transmitter using retrodirective array,''
\newblock {\em IEEE Journal of Selected Topics in Signal Processing}, 2017.

\bibitem{ding2017Circular}
Y. Ding, V. Fusco, and A. Chepala,
\newblock ``Circular directional modulation transmitter array,''
\newblock {\em {IET} Microwaves, Antennas Propagation}, 2017.

\bibitem{Alotaibi2016subset}
N.~N. Alotaibi and K.~A. Hamdi,
\newblock ``Switched phased-array transmission architecture for secure millimeter-wave wireless communication,''
\newblock {\em IEEE Transactions on Communications}, vol. 64, no. 3, pp. 1303--1312, 2016.

\bibitem{Ding2019TMA_OFDM}
Y. Ding, V. Fusco, J. Zhang, and W.-Q. Wang,
\newblock ``Time-modulated {OFDM} directional modulation transmitters,''
\newblock {\em IEEE Transactions on Vehicular Technology}, 2019.

\bibitem{nooraiepour2022time}
A. Nooraiepour, S. Vosoughitabar, C.-T.~M. Wu, W.~U. Bajwa, and N.~B. Mandayam,
\newblock ``Time-varying metamaterial-enabled directional modulation schemes for physical layer security in wireless communication links,''
\newblock {\em ACM Journal on Emerging Technologies in Computing Systems}, vol. 18, no. 4, pp. 1--20, 2022.

\bibitem{Purushothama2023Synthesis}
J.~M. Purushothama, Y. Ding, G. Goussetis, G. Huang, and Y. Xiao,
\newblock ``Synthesis of energy efficiency-enhanced directional modulation transmitters,''
\newblock {\em IEEE Transactions on Green Communications and Networking}, 2023.

\bibitem{Guo2018time}
J. Guo, L. Poli, M.~A. Hannan, P. Rocca, S. Yang, and A. Massa,
\newblock ``Time-modulated arrays for physical layer secure communications: {O}ptimization-based synthesis and experimental assessment,''
\newblock {\em IEEE Transactions on Antennas and Propagation}, 2018.

\bibitem{li2022chaotic}
H. Li, Y. Chen, and S. Yang,
\newblock ``Chaotic-enabled phase modulation in time-modulated arrays for secure transmission,''
\newblock {\em IEEE Transactions on Antennas and Propagation}, vol. 70, no. 11, pp. 10454--10464, 2022.

\bibitem{Huang2021multi}
G. Huang, Y. Ding, and S. Ouyang,
\newblock ``Multicarrier directional modulation symbol synthesis using time-modulated phased arrays,''
\newblock {\em IEEE Antennas and Wireless Propagation Letters}, vol. 20, no. 4, pp. 567--571, 2021.

\bibitem{Huang2022target}
G. Huang, Y. Ding, S. Ouyang, and J.~M. Purushothama,
\newblock ``Target localization using time-modulated directional modulated transmitters,''
\newblock {\em IEEE Sensors Journal}, 2022.

\bibitem{xu2023TMA}
Z. Xu and A.~P. Petropulu,
\newblock ``A secure dual-function radar communication system via time-modulated arrays,''
\newblock in {\em 2023 IEEE Radar Conference (RadarConf23)}, 2023.

\bibitem{Aapo2000ica}
A. Hyv\"{a}rinen and E. Oja,
\newblock ``Independent component analysis: {A}lgorithms and applications,''
\newblock {\em Neural Networks}, vol. 13, no. 4-5, pp. 411--430, 2000.

\bibitem{Aapo1999fast}
A. Hyv\"{a}rinen,
\newblock ``Fast and robust fixed-point algorithms for independent component analysis,''
\newblock {\em IEEE Transactions on Neural Networks}, 1999.

\end{thebibliography}
\bibliographystyle{IEEEbib_abbr_xu}

\end{document}